%% file: main_arXiv.tex
\documentclass[twocolumn,prb,preprintnumbers,amsmath,amssymb,superscriptaddress,letterpaper]{revtex4-2}

\usepackage[dvipdfmx]{graphicx}
\usepackage{color}
\usepackage[normalem]{ulem}
\usepackage{newfloat}
\usepackage{amsmath}
\usepackage{setspace}
\usepackage{lineno}
\usepackage{bm}
\usepackage{float}
\DeclareFloatingEnvironment[name={Supplementary Figure S}]{suppfigure}

\setcitestyle{super,sort&compress,open={},close={}}

\usepackage[dvipdfmx, colorlinks=true, citecolor=blue, linkcolor=blue, urlcolor=blue]{hyperref}

\begin{document}
\title{Realization of a parity-violating antiferromagnetic state in LaMnSi}
	\author{Takuma Iwata}
	\affiliation{Graduate School of Advanced Science and Engineering, Hiroshima University, Higashi-Hiroshima 739-8526, Japan}
	\affiliation{International Institute for Sustainability with Knotted Chiral Meta Matter (WPI-SKCM$^2$), Hiroshima University, Higashi-Hiroshima, 739-8531, Japan}
	
	\author{K.~Shiraishi} 
	\affiliation{Graduate School of Advanced Science and Engineering, Hiroshima University, Higashi-Hiroshima 739-8526, Japan}
	
	\author{T.~Aoyama} 
	\affiliation{Graduate School of Advanced Science and Engineering, Hiroshima University, Higashi-Hiroshima 739-8526, Japan}
	\affiliation{International Institute for Sustainability with Knotted Chiral Meta Matter (WPI-SKCM$^2$), Hiroshima University, Higashi-Hiroshima, 739-8531, Japan}
	
	\author{D.~Senba} 
	\affiliation{Graduate School of Advanced Science and Engineering, Hiroshima University, Higashi-Hiroshima 739-8526, Japan}
	
	\author{T.~Takeda}
    \affiliation{Graduate School of Advanced Science and Engineering, Hiroshima University, Higashi-Hiroshima 739-8526, Japan}
	\affiliation{Department of Chemical System Engineering, Graduate School of Engineering, The University of Tokyo, Bunkyo-ku, Tokyo 113-8656, Japan}
	
	\author{Y.~Fujisawa} 
	\affiliation{Research Institute for Synchrotron Radiation Science, Hiroshima University, Higashi-Hiroshima 739-0046, Japan}
	
	\author{M.~Nurmamat} 
	\affiliation{Graduate School of Advanced Science and Engineering, Hiroshima University, Higashi-Hiroshima 739-8526, Japan}
	
	\author{K.~Nakanishi} 
    \affiliation{Graduate School of Advanced Science and Engineering, Hiroshima University, Higashi-Hiroshima 739-8526, Japan}
	
	\author{K.~Yamagami} 
	\affiliation{Japan Synchrotron Radiation Research Institute (JASRI), Sayo, Hyogo 679-5198, Japan}

    \author{M.~Arita} 
	\affiliation{Research Institute for Synchrotron Radiation Science, Hiroshima University, Higashi-Hiroshima 739-0046, Japan}

    \author{T.~Yamada} 
	\affiliation{Liberal Arts and Sciences, Toyama Prefectural University, Imizu, Toyama 939-0398, Japan}
    
	\author{Y.~Yanagi} 
	\affiliation{Liberal Arts and Sciences, Toyama Prefectural University, Imizu, Toyama 939-0398, Japan}
	
	\author{A.~Kimura} 
	\affiliation{Graduate School of Advanced Science and Engineering, Hiroshima University, Higashi-Hiroshima 739-8526, Japan}
	\affiliation{International Institute for Sustainability with Knotted Chiral Meta Matter (WPI-SKCM$^2$), Hiroshima University, Higashi-Hiroshima, 739-8531, Japan}
	\affiliation{Research Institute for Semiconductor Engineering, Hiroshima University, Higashi-Hiroshima 739-8527, Japan}
	\affiliation{Synchrotron Radiation Research Center, National Institutes for Quantum Science and Technology (QST), Sayo-gun, Hyogo 679-5148, Japan}
	
	\author{H.~Tanida} 
	\affiliation{Liberal Arts and Sciences, Toyama Prefectural University, Imizu, Toyama 939-0398, Japan}
	
	\author{Kenta~Kuroda}
	\affiliation{Graduate School of Advanced Science and Engineering, Hiroshima University, Higashi-Hiroshima 739-8526, Japan}
	\affiliation{International Institute for Sustainability with Knotted Chiral Meta Matter (WPI-SKCM$^2$), Hiroshima University, Higashi-Hiroshima, 739-8531, Japan}
	\affiliation{Research Institute for Semiconductor Engineering, Hiroshima University, Higashi-Hiroshima 739-8527, Japan} 
\date{\today}
\maketitle
\section*{\protect\NoCaseChange{Abstract}}
\textbf{
Spontaneous symmetry breaking underlies functional electronic phenomena in quantum materials.
Breaking space-inversion ($\mathcal{P}$) or time-reversal ($\mathcal{T}$) symmetry can generate spin-split electronic bands central to modern spintronics.
By contrast, parity-violating antiferromagnetic (AFM) order breaks both $\mathcal{P}$ and $\mathcal{T}$ while preserving the combined $\mathcal{PT}$ symmetry, enabling spin-degenerate yet momentum-asymmetric electronic bands.
This momentum asymmetry has been proposed as a microscopic origin of unconventional nonreciprocal and nonlinear responses but its experimental verification has remained challenging because it requires establishing both the symmetry-breaking magnetic order and the associated electronic structure.
Here we combine soft x-ray angle-resolved photoemission spectroscopy (ARPES) and polarization-resolved optical second-harmonic generation (SHG) microscopy to study LaMnSi, a candidate parity-violating AFM metal.
Soft x-ray ARPES resolves the three-dimensional bulk band structures in agreement with density functional theory calculations for the AFM phase, whereas SHG microscopy detects sign-reversing nonlinear optical responses from opposite AFM domains that carry $\mathcal{T}$-odd parity-violating order.
Together, these results provide direct evidence for parity-violating AFM state in LaMnSi, establish LaMnSi as a parity-violating AFM metal, and identify this class of AFMs as a promising platform for symmetry-controlled nonreciprocal and nonlinear electronic responses.
}

\begin{figure*}[t!]
\begin{center}
\includegraphics{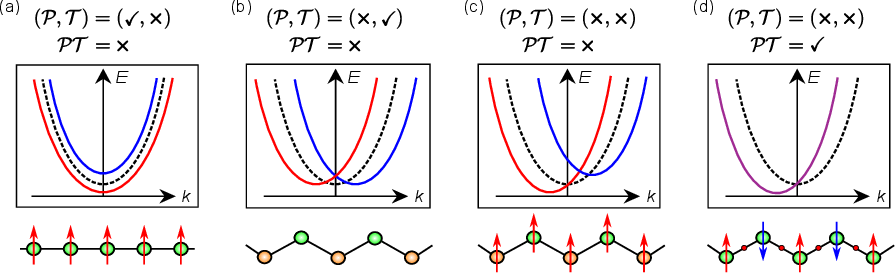}
\caption{
\textbf{Classification of electronic band structures based on $\mathcal{P}$, $\mathcal{T}$ and $\mathcal{PT}$ symmetries.}
(\textbf{a}) When $\mathcal{P}$ is preserved while $\mathcal{T}$ is broken, exchange spin splitting occurs but the band dispersion remains momentum symmetric. 
(\textbf{b}) When $\mathcal{P}$ is broken while $\mathcal{T}$ is preserved, Rashba-type spin splitting appears. 
(\textbf{c}) When both $\mathcal{P}$ and $\mathcal{T}$ are broken without symmetry constraint, spin-split bands can exhibit momentum asymmetry. 
(\textbf{d}) In contrast, the parity-violating AFM order studied in this work represents a special case in which both $\mathcal{P}$ and $\mathcal{T}$ are broken in a correlated manner while the composite $\mathcal{PT}$ symmetry is preserved, enforcing spin degeneracy while allowing momentum-asymmetric band dispersions. 
Red and blue curves represent spin-split bands, whereas the purple curve denotes spin-degenerate asymmetric bands; dashed curves indicate reference bands with both $\mathcal{P}$ and $\mathcal{T}$ preserved.
In the bottom panels, the zigzag chains illustrate a minimal lattice model used to visualize and compare the symmetry conditions in panels (\textbf{a})-(\textbf{d}). 
In this model as shown in (\textbf{d}), the $\mathcal{P}$ operation exchanges the two sublattices, while the subsequent $\mathcal{T}$ operation reverses the magnetic orientation, so that the combined $\mathcal{PT}$ operation maps each state onto itself, providing a simple realization of parity-violating AFM order. 
The large circles represent atomic sites and arrows denote magnetic moments. 
The small circles in (\textbf{d}) indicate the crystallographic $\mathcal{P}$ center, which is broken by the AFM order.
}
\label{fig1}
\end{center}
\end{figure*}
\begin{figure*}[t!]
\begin{center}
\includegraphics{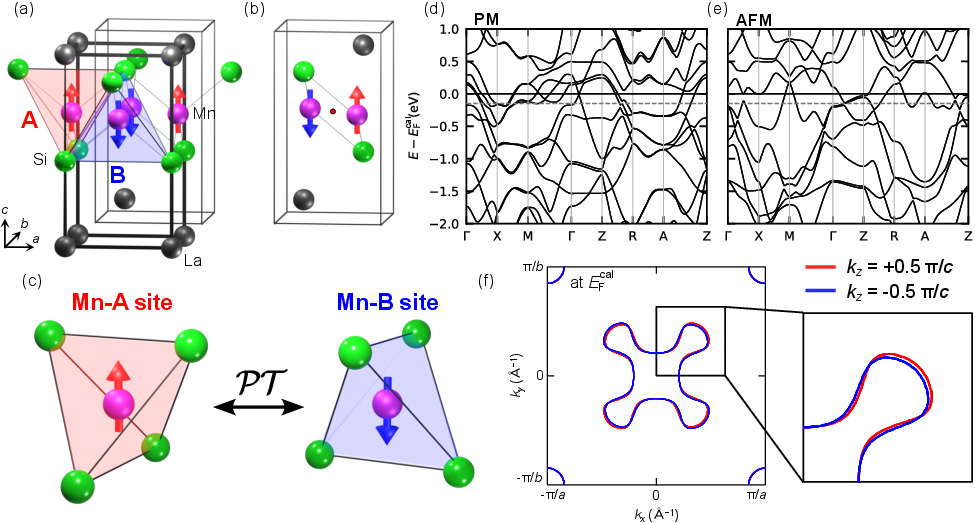}
\caption{
\textbf{Magnetic structure and calculated three-dimensional bulk electronic structure of LaMnSi.} 
(\textbf{a}) AFM structure of LaMnSi in the nonsymmorphic lattice. 
The Mn moments form a collinear AFM order and align along the [001] direction and with a zero propagation vector ($\bm{q}$=$\bm{0}$), as indicated by red and blue arrows. 
Thick lines denote the primitive unit cell. 
(\textbf{b}) Alternative unit cell [thin line in (\textbf{a})], where the small red circle indicates the crystallographic $\mathcal{P}$ center, which is no longer a symmetry of the magnetic structure in the AFM order.
(\textbf{c}) Red and blue tetrahedra isolated from (\textbf{a}), representing the two Mn sublattices (A and B).
In the absence of magnetic moment, the two sublattices are related by inversion about the center indicated by the solid circle in (\textbf{b}). 
When parity-violating AFM order sets in, this inversion equivalence is lifted, and the sublattices are no longer related by $\mathcal{P}$ or $\mathcal{T}$ individually, but instead by the combined $\mathcal{PT}$ operation.
(\textbf{d}), (\textbf{e}) Calculated band structures with spin–orbit coupling along high-symmetry lines for the paramagnetic (PM) and AFM phases. 
The solid line indicates the calculated Fermi level ($E_\mathrm{F}^{\mathrm{cal}}$), while the dashed line marks the experimental Fermi level ($E_\mathrm{F}$) determined by ARPES. 
An energy offset (0.15~eV) is applied to facilitate comparison between the DFT calculation and the ARPES results (see Fig.~\ref{fig3}). 
(\textbf{f}) Constant-energy contours at $E_\mathrm{F}^{\mathrm{cal}}$ in the AFM phase for $k_z=+0.5\pi/c$ (red) and $k_z=-0.5\pi/c$ (blue). 
Inset: a magnified view of the $k_x$–$k_y$ region, highlighting the spin-degenerate momentum-space asymmetry [$E(k_x, k_y, k_z) \neq E(k_x, k_y, -k_z)$], which is consistent with the parity-violating AFM state illustrated in Fig.~\ref{fig1}(\textbf{d}) (see also Supplementary Note~1 for more details).
}
\label{fig2}
\end{center}
\end{figure*}
\begin{figure*}[thb]
\begin{center}
\includegraphics{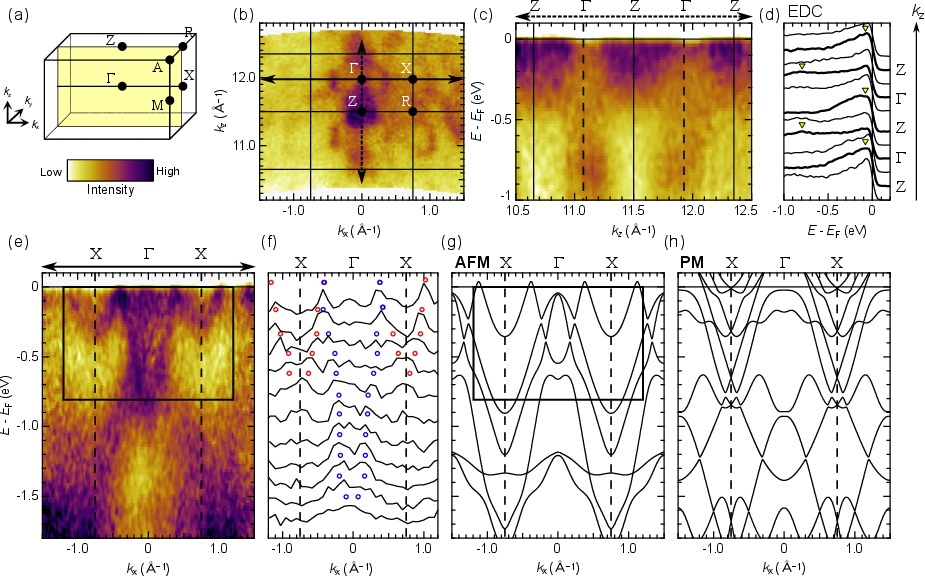}
\caption{
\textbf{Three-dimensional electronic structure in the AFM phase.}
(\textbf{a}) Brillouin zone with high-symmetry points and the definition of $k_{x,y,z}$.
(\textbf{b}) Fermi-surface map in the $k_x$–$k_z$ plane [yellow plane in (\textbf{a})], obtained from photon-energy-dependent SX-ARPES measurements with $h\nu$=400--600~eV at 50~K, well below $T_{\rm{N}}$. 
(\textbf{c}) Band dispersion along the $\Gamma$–Z direction [dashed arrow in (\textbf{b})], revealing a clear $k_z$-dependent modulation. 
(\textbf{d}) Energy distribution curves (EDCs) at representative $k_z$ values extracted from (\textbf{c}). 
Bold curves highlight spectra at the high-symmetry $\Gamma$ and Z points. 
The peak positions (yellow triangles) exhibit a periodic variation consistent with the bulk Brillouin zone along $k_z$. 
(\textbf{e}) ARPES intensity map along the $\Gamma$–X direction [solid arrow in (\textbf{b})]. 
(\textbf{f}) Momentum distribution curves (MDCs) extracted from the energy–momentum window indicated in (\textbf{e}).
The band dispersion is traced by the MDC peak positions (circles). 
(\textbf{g}, \textbf{h}) Calculated band structures along $\Gamma$–X for the AFM and PM phases. 
To facilitate comparison with experiment, the calculated bands are rigidly shifted to lower energy by $0.15$~eV [see Figs.~\ref{fig2}(\textbf{d}, \textbf{e}) and Supplementary Note~3].
}			
\label{fig3}
\end{center}
\end{figure*}
\begin{figure*}[thb]
\begin{center}
\includegraphics{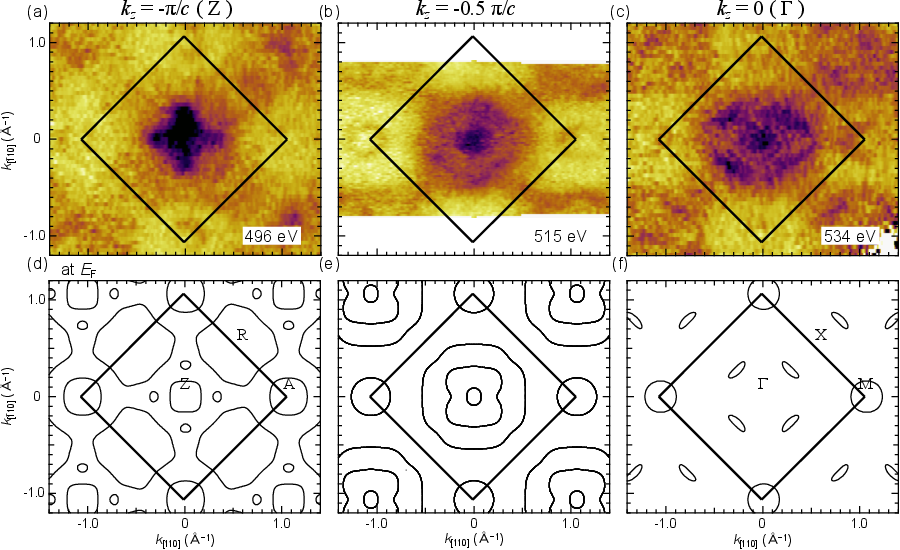}
\caption{
\textbf{Fermi-surface mappings at selected $k_z$ values.}
(\textbf{a}–\textbf{c}) ARPES intensity maps at $E_\mathrm{F}$ measured at $k_z \approx -\pi/c$ (\textbf{a}), $-0.5\pi/c$ (\textbf{b}), and $0$ (\textbf{c}), using photon energies ($h\nu$) of 496, 515, and 534~eV, respectively. 
The solid squares indicate the projected tetragonal Brillouin zone [see Fig.~\ref{fig3}(\textbf{a})]. 
(\textbf{d}–\textbf{f}) Corresponding calculated constant-energy contours at $E_\mathrm{F}$ for the same $k_z$ planes. 
The calculated bands are rigidly shifted to lower energy by $0.15$~eV [see Figs.~\ref{fig2}(\textbf{d}, \textbf{e}) and Supplementary Note~3].
}			
\label{fig4}
\end{center}
\end{figure*}

%
%
\begin{figure*}[htbp]
\begin{center}
\includegraphics{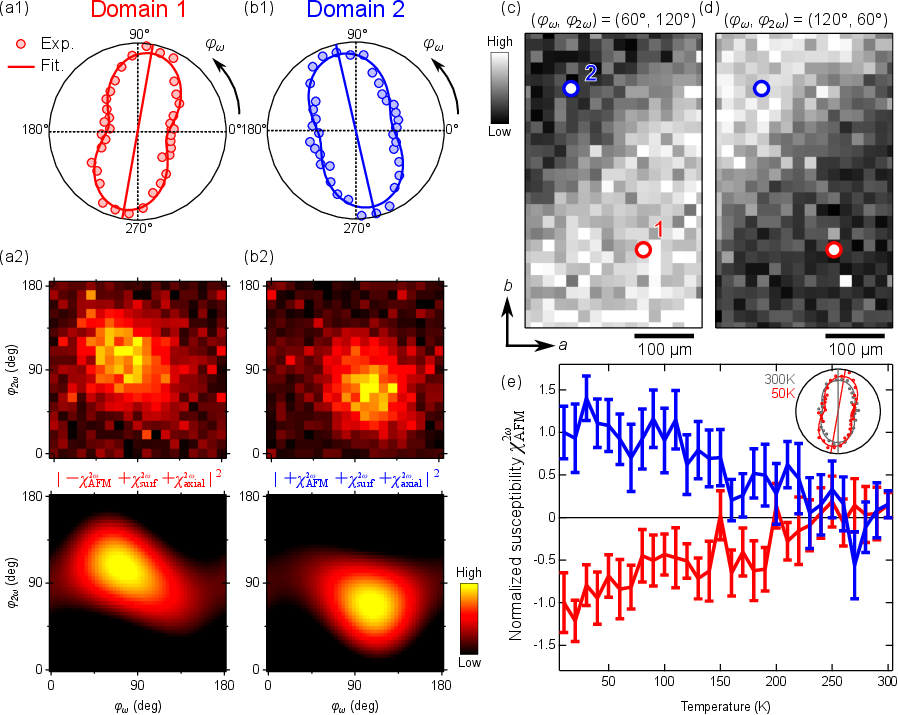}
\caption{
\textbf{Polarization-resolved SHG microscopy and temperature evolution of the nonlinear susceptibility.}
(\textbf{a1}, \textbf{b1}) Polar plots of the SHG intensity as a function of the fundamental polarization angle ($\varphi_\omega$) at two spatial locations corresponding to opposite AFM configurations (domain~1 and domain~2) on the (001) surface at 15~K (see Supplementary Note~6 for experimental setup).
The SHG polarization angle $\varphi_{2\omega}$ is fixed at $90^\circ$.
Solid curves are fits based on the nonlinear susceptibility model (Supplementary Note~7).
The opposite angular shifts of the intensity maxima directly reflect the sign reversal of the $\mathcal{T}$-odd nonlinear susceptibility $\chi^{2\omega}_{\mathrm{AFM}}$.
(\textbf{a2}, \textbf{b2}) Top: two-dimensional SHG intensity maps as functions of $\varphi_\omega$ and $\varphi_{2\omega}$ measured at the same locations as in (\textbf{a1}) and (\textbf{b1}).
Bottom: corresponding maps reproduced by global fitting based on the magnetic point-group symmetry, enabling quantitative separation of the nonlinear susceptibility components: the AFM-induced $\mathcal{T}$-odd component $\chi^{2\omega}_{\mathrm{AFM}}$, the crystallographic and surface polar components $\chi^{2\omega}_{\mathrm{axial}}$ and $\chi^{2\omega}_{\mathrm{surf}}$ (Supplementary Note~7). 
(\textbf{c}, \textbf{d}) Scanning SHG microscopy images acquired under polarization configurations that maximize the contrast: $(\varphi_\omega, \varphi_{2\omega}) = (60^\circ, 120^\circ)$ (\textbf{c}) and $(120^\circ, 60^\circ)$ (\textbf{d}).
The contrast reversal directly visualizes the sign change of $\chi^{2\omega}_{\mathrm{AFM}}$ between the parity-violating AFM domains.
Scale bar, 100~\textmu m.
(\textbf{e}) Temperature dependence of the $\mathcal{T}$-odd $\chi^{2\omega}_{\mathrm{AFM}}$ for the two AFM configurations (red and blue), normalized to their respective values at 10~K.
The signal emerges below $\sim200$~K and increases upon cooling with opposite signs, despite the Néel temperature being $T_{\rm N}=293$~K.
Error bars denote the standard error of the fitting.
Inset: representative polar plots at 50~K and 300~K.
}
\label{fig5}
\end{center}
\end{figure*}
\section*{\protect\NoCaseChange{Introduction}}
Spontaneous symmetry breaking is a fundamental organizing principle of condensed matter physics, linking microscopic electronic structures to emergent macroscopic responses in solids~\cite{tokura2017nphys}. 
Among these responses, nonreciprocal electronic phenomena, in which transport properties depend on the propagation direction, have become a central theme in modern quantum materials research~\cite{tokura2018nonreciprocal,Nadeem_nrevphys2023}.  
Crucially, the simultaneous breaking of space-inversion ($\mathcal{P}$) and time-reversal ($\mathcal{T}$) symmetries induces such direction-dependent responses~\cite{nagaosa2024review}.
Microscopically, their origins include momentum asymmetry in electronic band structures and asymmetric scattering processes~\cite{Rikken_prl2001}.

In general, breaking either $\mathcal{P}$ or $\mathcal{T}$ symmetry produces spin-split electronic structures~\cite{Okuda2013, Dil2019} [Figs.~\ref{fig1}({\bf a}, {\bf b})], which have been widely studied as platforms for spin-dependent phenomena~\cite{Manchon_nmat2015} and spintronics~\cite{rmp2004}.
When both $\mathcal{P}$ and $\mathcal{T}$ symmetries are broken simultaneously, the system develops a  momentum asymmetry in its electronic band structure~\cite{Ideue_nphys2017} [Fig.~\ref{fig1}({\bf c})], leading to nonreciprocal responses. 
Experimentally, most realizations of nonreciprocal electronic responses have relied on noncentrosymmetric crystals or interfaces, where $\mathcal{P}$ symmetry is intrinsically broken while $\mathcal{T}$ symmetry is externally broken by applying strong magnetic fields~\cite{Ideue_nphys2017, watatsuki2017sciadv, choe2019ncom, Kondo2025prr, he2019ncom}.

However, antiferromagnetic (AFM) state in metals has recently emerged as a fundamentally different and powerful route for controlling the electronic band structure and the resulting emergent quantum properties~\cite{Zelenzny2014prl, Yuan2021prm, hayami2019jpsj, hayami2024jpsj, watanabe2017prb, jungwirth2026review}.
In this context, a distinct and important case is that of parity-violating AFM states in $\mathcal{PT}$-symmetric systems.
Such states arise in crystals with sublattice degrees of freedom that locally lack $\mathcal{P}$ symmetry, even though the crystal as a whole remains centrosymmetric~\cite{yanase2014}.
When AFM order develops on these sublattices, the hidden local inversion asymmetry is activated, giving rise to a magnetic order that breaks both $\mathcal{P}$ and $\mathcal{T}$ simultaneously while preserving the combined $\mathcal{PT}$ symmetry~\cite{watanabe2021prx, Winkler_prb2023, Watanabe2024}.
This symmetry setting is fundamentally distinct from altermagnetism, where compensated magnetic order leads to spin-split bands through breaking $\mathcal{PT}$ symmetry~\cite{jungwirth2026review}.
Instead, parity-violating AFM states host spin-degenerate yet momentum-asymmetric electronic states [$E(k)\neq E(-k)$, Fig.~\ref{fig1}({\bf d})], which underpin nonreciprocal responses~\cite{wang2023nature, Gao2023science, Yang2019ncom, sun2019nature, Mayo2025pnas}.
The essential symmetry principle is captured by a minimal zigzag-chain model with collinear AFM order~\cite{yanase2014} [see also the bottom of Fig.~\ref{fig1}({\bf d})].
A hallmark of this state is that spontaneous $\mathcal{P}$-breaking has a $\mathcal{T}$-odd character. 
Accordingly, the $\mathcal{T}$ operation, which relates AFM domains with opposite spin configurations, also reverses the AFM-induced $\mathcal{P}$-breaking.
As a result, both the associated momentum-space asymmetry and the corresponding responses reverse sign between opposite AFM domains.

To date, parity-violating AFM states have been theoretically predicted in a number of magnetic materials based on group-theoretical classification~\cite{Yuan2021prm,watanabe2018prb}. 
Despite this theoretical progress, experimental verification of electronic states characterized by momentum-asymmetric bands has so far been limited.
Recent nonreciprocal transport measurements have reported signatures consistent with such symmetry breaking in several candidate systems~\cite{kimata2026prl, sakai2025ncom}, yet establishing a direct link between the symmetry-breaking magnetic order and the associated electronic state remains challenging.

Angle-resolved photoemission spectroscopy (ARPES)~\cite{sobota_rmp2021}, including its spin-resolved implementation~\cite{Okuda2013, Dil2019}, provides a powerful probe that directly accesses electronic band structures. 
Using this technique, exchange spin splitting in ferromagnets~\cite{ye2018nature} [Fig.~\ref{fig1}({\bf a})], Rashba-type spin splitting~\cite{lashell1996prl} [Fig.~\ref{fig1}({\bf b})], and asymmetric spin splitting arising from the simultaneous breaking of $\mathcal{P}$ and $\mathcal{T}$ symmetries~\cite{krempasky2016ncom, Grenz_njp2023, edwards2023nmat} [Fig.~\ref{fig1}({\bf c})] have been experimentally identified.
In all of these cases, the characterisstic signatures appear through spin splitting of electronic bands.

In contrast, parity-violating AFM states remain spin degenerate exhibiting momentum-asymmetric band sturctures~\cite{yanase2014} [Fig.~\ref{fig1}({\bf d})]. 
Detecting such a momentum asymmetry is intrinsically challenging because it appears without spin splitting and is further obscured by averaging over AFM domains.
Although spatially resolved ARPES has reported photoemission-intensity asymmetry consistent with magnetic parity violation~\cite{Fedchenko2022jpcm,Lytvynenko2023prb}, establishing its $\mathcal{T}$-odd character remains nontrivial and requires an independent probe sensitive to the reversal of the momentum-space asymmetry between parity-breaking AFM domains.

Here, we address this challenge by combining soft x-ray ARPES (SX-ARPES) and polarization-resolved optical second-harmonic generation (SHG) microscopy on LaMnSi, a candidate parity-violating AFM metal.
SX-ARPES resolves the three-dimensional bulk electronic structure in good agreement with density functional theory (DFT) calculations for the AFM phase.
Complementarily, the $\mathcal{T}$-odd character of the parity-violating AFM state is directly identified by polarization-resolved SHG microscopy through a nonlinear susceptibility whose sign reverses between opposite AFM domains.
The temperature evolution of this nonlinear susceptibility further reveals the gradual development of the $\mathcal{T}$-odd response upon cooling, highlighting its intimate connection to the parity-violating AFM order.
These results identify LaMnSi as a parity-violating AFM metal and establish a direct route toward accessing symmetry-controlled nonlinear responses rooted in momentum-asymmetric electronic states.

\section*{\protect\NoCaseChange{Results}}
\noindent\textbf{AFM order in the nonsymmorphic lattice of LaMnSi.}
We first describe the crystal and magnetic structures of LaMnSi~\cite{tanida2022jpsj}. 
As shown in Fig.~\ref{fig2}(\textbf{a}), LaMnSi crystallizes in the nonsymmorphic space group $P4/nmm$, whose unit cell contains two Mn sublattices (highlighted by red and blue tetrahedra). 
Although the crystal as a whole preserves $\mathcal{P}$ symmetry (crystal class $D_{4h}$), it is locally broken at individual Mn sites (site symmetry $D_{2d}$). 
Below the Néel temperature ($T_{\rm N}=293$~K), the Mn moments form a collinear AFM order and align along the $c$ axis~\cite{tanida2022jpsj, sakai2026jpsj}. 
In the absence of magnetic moments, the two sublattices can be exchanged by the $\mathcal{P}$ operation. 
Once the AFM order develops, however, this $\mathcal{P}$ operation no longer preserves the magnetic configuration [Fig.~\ref{fig2}(\textbf{b})], and consequently the AFM order breaks $\mathcal{P}$ symmetry in addition to $\mathcal{T}$ symmetry.
Nevertheless, the original configuration can be recovered by a subsequent $\mathcal{T}$ operation, so that the combined $\mathcal{PT}$ symmetry remains preserved [Fig.~\ref{fig2}(\textbf{c})], analogous to the zigzag model~\cite{yanase2014} [see Fig.~\ref{fig1}(\textbf{d})].
As a result, two distinct AFM configurations in the nonsymmorphic sublattices, related by the $\mathcal{T}$ operation, exist and form degenerate domains. 
These symmetry relations indicate that the AFM structure of LaMnSi satisfies the symmetry conditions for a parity-violating AFM state~\cite{Watanabe2024}.\\

\noindent\textbf{Predicted asymmetric electronic structures.}
To examine how this symmetry setting manifests in the electronic structure, we calculated the band dispersions including spin–orbit coupling (SOC) for both the paramagnetic (PM) and AFM phases [Figs.~\ref{fig2}(\textbf{d}, \textbf{e})]. 
Although the AFM structure has a zero propagation vector~\cite{tanida2022jpsj, Welter1994} ($\bm{q}=\bm{0}$), the electronic structure undergoes a pronounced reconstruction in the AFM phase. 
This reconstruction associated with the AFM order originates from the combined effects of exchange interactions and the breaking of the nonsymmorphic crystal symmetry. 

As shown in Fig.~\ref{fig2}(\textbf{f}), the calculated constant-energy contours at out-of-plane momenta $k_z=\pm0.5\pi/c$ are no longer identical: the Fermi contours exhibit a clear asymmetry between positive and negative $k_z$. 
This directly implies $E(k_x,k_y,k_z)\neq E(k_x,k_y,-k_z)$, providing a characteristic momentum-space signature of the parity-violating AFM state.
Beyond this $k_z$ asymmetry, each constant-$k_z$ contour also exhibits pronounced anisotropy in the $k_x$–$k_y$ plane.
We find that these asymmetric features are well captured by a $k_xk_yk_z$ term from the perspective of magnetic point group~\cite{watanabe2017prb} and emerge only when SOC is included in the calculations (see Supplementary Note~1).
Because the band asymmetry is odd under the $\mathcal{T}$ operation, its sign reverses between the two domains with opposite $\mathcal{T}$ configurations.  
Consequently, if such domains coexist, their contributions average out in momentum space, making the intrinsic band asymmetry difficult to determine experimentally.\\

\noindent\textbf{Bulk electronic structure experimentally revealed by SX-ARPES.}
The three-dimensional electronic structure was investigated using synchrotron SX-ARPES.
The use of soft x-rays enhances the photoelectron escape depth, providing increased bulk sensitivity and $k_z$ resolution~\cite{Strocov2014jsr, kuroda_prl2018}.
These measurements were performed at 50~K, well below $T_{\rm N}$.
Figures~\ref{fig3}(\textbf{b}) and~\ref{fig3}(\textbf{c}) show the photoelectron intensity map at the Fermi energy ($E_{\rm{F}}$) in the $k_x$–$k_z$ momentum plane [the yellow plane in Fig.~\ref{fig3}(\textbf{a})] and the band map along the $\Gamma$–Z direction [along the dashed line in Fig.~\ref{fig3}(\textbf{b})], respectively. 
A clear $k_z$-periodic modulation of the photoelectron intensity is observed in both measurements. 
This periodic modulation is further confirmed by the energy distribution curves (EDCs) shown in Fig.~\ref{fig3}(\textbf{d}). 
The spectral weight approaches $E_{\rm{F}}$ at the Z points, whereas it is observed at $E-E_{\rm{F}}=-0.8$~eV at the $\Gamma$ points, repeating periodically along the $k_z$ direction (yellow markers). 
This behavior follows the periodicity of the bulk Brillouin zone along $k_z$, confirming the bulk origin of the observed electronic structure.

Based on the clear $k_z$ periodicity, we identify the photon energy corresponding to the $\Gamma$ point and obtain the ARPES intensity map along the high-symmetry $\Gamma$–X direction, shown in Fig.~\ref{fig3}(\textbf{e}). 
We observe an electron-like band centered at the $\Gamma$ point together with another centered at the X point. 
The corresponding dispersions are traced by the peak positions of the momentum distribution curves (MDCs) in Fig.~\ref{fig3}(\textbf{f}), forming electron-like bands with minima located at the respective high-symmetry points. 
While the MDC analysis reveals the electron-like dispersions, the band map in Fig.~\ref{fig3}(\textbf{e}) further shows that the electron-like band centered at the $\Gamma$ point appears to intersect with a hole-like band around $E-E_{\rm{F}}\sim{-}$0.5~eV.

We note that no significant changes in the band structures were observed upon increasing the temperature well above $T_{\rm{N}}$ even up to 320~K (Supplementary Note~2).
To gain insight into the observed dispersions, we therefore compare the experimental results with the theoretical band structures for the PM and AFM phases [Figs.~\ref{fig3}(\textbf{g}, \textbf{h})].
The AFM calculation reproduces the electron pocket at the X point and captures the overall dispersion near the $\Gamma$ point, including the crossing between the electron-like and hole-like bands. 
In contrast, these characteristic dispersive features are not reproduced in the PM calculation.
The better agreement with the AFM calculation suggests that the AFM order influences the bulk electronic structure in LaMnSi.
The absence of pronounced temperature evolution in our ARPES results indicates that the exchange splitting of Mn $3d$ orbitals remains robust even above $T_{\rm N}$, implying that the electronic structure is likely governed by short-range magnetic interactions rather than long-range order.

We further examined the momentum-space distribution of electronic states near $E_{\rm F}$ at several photon energies corresponding to different $k_z$ planes using SX-ARPES. 
Figures~\ref{fig4}(\textbf{a})–(\textbf{c}) show the resulting FS maps acquired with different photon energies, revealing systematic changes in the photoelectron distribution. 
At the $\Gamma$ plane, the intensity distribution exhibits four characteristic elliptical features surrounding the zone center [Fig.~\ref{fig4}(\textbf{c})]. 
In contrast, in the Z plane these features are absent, and a cross-shaped intensity pattern appears at the center [Fig.~\ref{fig4}(\textbf{a})]. 
Such systematic evolution of the intensity distribution with $k_z$ is well reproduced by the calculated constant-energy contours for the AFM phase [Figs.~\ref{fig4}(\textbf{d})–(\textbf{f})].

We also note that pronounced photoelectron intensity is observed around the $\Gamma$ point in the experimental map [Fig.~\ref{fig4}(\textbf{c})], although the calculation predicts no pockets at this momentum [Fig.~\ref{fig4}(\textbf{f})].
This mismatch likely originates from the spectral tail of the bands just below $E_{\rm F}$ due to finite experimental energy resolution and $k_z$ broadening (Supplementary Note~4).

In addition, we find that the observed electronic bands near $E_{\rm F}$ are dominated by Mn $3d$ orbitals hybridized with La $5d$ orbitals (Supplementary Note~5). 
This indicates that the Mn $3d$ electrons simultaneously form the magnetic order and act as the itinerant carriers. 
Such a dual role is particularly important in the present system, as it allows the itinerant electrons to directly couple to the  AFM order, providing a natural microscopic origin for the momentum-asymmetric electronic structure expected in this state.

Having established the bulk electronic structure of LaMnSi by SX-ARPES and its consistency with the theoretical calculations for the AFM phase, we next examine possible asymmetry in the electronic structure expected for this ordered state.
In LaMnSi, the momentum asymmetry is captured by a $k_x k_y k_z$ term within the magnetic point group~\cite{watanabe2017prb} (Supplementary Note 1), implying that it appears only away from high-symmetry $k_z$ planes [Fig.~\ref{fig4}(\textbf{e})].
Experimentally, resolving such asymmetry of the electronic structures and identifying its $\mathcal{T}$-odd character remain inherently challenging with ARPES measurements that rely on photoemission-intensity distributions~\cite{Fedchenko2022jpcm,Lytvynenko2023prb}.
Therefore, although our SX-ARPES establishes the bulk electronic-structure basis of the AFM phase in LaMnSi, accessing the parity-violating AFM state requires a complementary symmetry-sensitive probe.\\

\noindent\textbf{$\mathcal{T}$-odd nonlinear susceptibility revealed by SHG microscopy.}
A defining feature of the parity-violating AFM state is that the spontaneous $\mathcal{P}$ breaking carries a $\mathcal{T}$-odd character, leading to sign reversal of the associated nonlinear response between two AFM domains with opposite parity-violating order.
To directly identify this defining feature, we employ polarization-resolved SHG microscopy [see Methods and Supplementary Note~6 for the setup]. 
SHG is inherently sensitive to $\mathcal{P}$-breaking in the electronic states through the nonlinear susceptibility tensors~\cite {petersen2006nphys, sun2019nature, fiebig2005josab, wu2017nature}.
Because the AFM order in LaMnSi breaks the global $\mathcal{P}$ symmetry of the nonsymmorphic crystal lattice [Figs.~\ref{fig2}(\textbf{a})-(\textbf{c})], it activates $\mathcal{T}$-odd electric-dipole nonlinear susceptibility components associated with the parity-violating AFM order.
These components should be distinguished from axial magnetic SHG terms allowed in $\mathcal{P}$-even magnetic orders.
Consistent with the magnetic point group $4'/m'm'm$~\cite{birss1964symmetry}, the AFM-induced electric-dipole components ($\chi^{2\omega}_{\mathrm{AFM}}$) coexist with $\mathcal{T}$-even contributions, including crystallographic terms ($\chi^{2\omega}_{\mathrm{axial}}$) and surface polar terms ($\chi^{2\omega}_{\mathrm{surf}}$) (Supplementary Note~7).
The two parity-violating AFM domains with opposite $\mathcal{P}$-breaking order  are connected by the $\mathcal{T}$ operation, such that only $\chi^{2\omega}_{\mathrm{AFM}}$ reverses sign between the domains, whereas the $\mathcal{T}$-even components remain unchanged.
This domain-dependent sign reversal becomes observable through interference between $\chi^{2\omega}_{\mathrm{AFM}}$ and the $\mathcal{T}$-even components, thereby providing direct access to the $\mathcal{T}$-odd character of the parity-violating AFM state~\cite{sun2019nature,fiebig1994prl,fiebig2005josab}.
The resulting symmetry-enforced sign reversal serves as a direct experimental fingerprint of the parity-violating AFM state.

This interference manifests in the polarization-dependent SHG intensity patterns, enabling extraction of the sign-reversing $\chi^{2\omega}_{\mathrm{AFM}}$ component through polarization-resolved SHG measurements.
Figures~\ref{fig5}(\textbf{a1}) and (\textbf{b1}) show polar plots of the SHG intensity as a function of the fundamental polarization angle ($\varphi_\omega$) at two spatial positions on the sample surface at 15~K.  
The SHG response exhibits clear domain-dependent angular shifts: the intensity maximum is rotated clockwise from the high-symmetry configuration ($\varphi_\omega=90^\circ$) in domain~1 [Fig.~\ref{fig5}(\textbf{a1})], while it is rotated counterclockwise in domain~2 [Fig.~\ref{fig5}(\textbf{b1})].  
Such antisymmetric angular shifts of the SHG intensity maxima are explained by interference between the $\mathcal{T}$-odd $\chi^{2\omega}_{\mathrm{AFM}}$ and the $\mathcal{T}$-even susceptibility components [Supplementary Note~7].
Because only $\chi^{2\omega}_{\mathrm{AFM}}$ reverses sign under the $\mathcal{T}$ operation relating the opposite AFM domains, the opposite angular shifts provide direct evidence of a $\mathcal{T}$-odd nonlinear response against the background of the $\mathcal{T}$-even contributions of $\chi^{2\omega}_{\mathrm{axial}}$ and $\chi^{2\omega}_{\mathrm{surf}}$.

We further measure the SHG intensity as a function of both the fundamental and SHG polarization angles ($\varphi_\omega$, $\varphi_{2\omega}$) as presented in top of Figs.~\ref{fig5}(\textbf{a2}) and (\textbf{b2}).  
The resulting intensity maps exhibit antisymmetric patterns with respect to the high-symmetry configuration $(90^\circ,90^\circ)$ corresponding to the center of the panels, with opposite direction for the two AFM domains.  
Based on these polarization dependencies, we identify configurations that maximize the intensity contrast and perform scanning SHG microscopy [Figs.~\ref{fig5}(\textbf{c}, \textbf{d})].  
The resulting images exhibit a complete contrast reversal upon switching the polarization configuration.  
This contrast inversion directly visualizes the sign reversal of $\chi^{2\omega}_{\mathrm{AFM}}$ between the two parity-violating AFM domains, providing direct identification of the parity-violating AFM state.  

Moreover, the polarization-dependent intensity maps are quantitatively reproduced by global fitting based on the magnetic point-group symmetry [bottom of Figs.~\ref{fig5}(\textbf{a2}, \textbf{b2})], enabling extraction of all the nonlinear susceptibility tensors [Supplementary Note~8].  
Using the extracted tensor components, SHG microscopy further allows us to track the temperature evolution of $\chi^{2\omega}_{\mathrm{AFM}}$ [Fig.~\ref{fig5}(\textbf{e})].
The $\mathcal{T}$-odd $\chi^{2\omega}_{\mathrm{AFM}}$ is observable only below $\sim200$~K and enhances continuously upon cooling, with opposite signs for the two AFM domains.  
Remarkably, although the parity-violating AFM order sets at the N\'eel temperature ($T_{\rm N}=293$~K), $\chi^{2\omega}_{\mathrm{AFM}}$ remains negligible just below $T_{\rm N}$ and becomes pronounced only well below $\sim200$~K.
A similar delayed onset has been reported for the magnetopiezoelectric response in EuMnBi$_2$, where it coincides with a crossover from incoherent to coherent electronic transport~\cite{shiomi2019prl}. 
In LaMnSi, we find that the growth of $\chi^{2\omega}_{\mathrm{AFM}}$ at low temperatures correlates with the reported drop in electrical resistivity around 200~K~\cite{tanida2022jpsj}, indicating the development of coherent itinerant electronic states well below the AFM transition.  
These observations suggest that the parity-violating AFM states and the associated electronic response emerge when coherent itinerant electronic states are established against the background AFM order.

\section*{\protect\NoCaseChange{Conclusion}}
In summary, we establish a parity-violating AFM state in the metal LaMnSi through a combination of SX-ARPES and polarization-resolved SHG microscopy.
In particular, we directly resolve the $\mathcal{T}$-odd nonlinear optical susceptibility $\chi^{2\omega}_{\mathrm{AFM}}$ and visualize its domain-dependent sign reversal.  
This enables an unambiguous identification of the parity-violating AFM state from both real-space imaging and symmetry-resolved responses together with momentum-space information obtained by ARPES.  
Furthermore, the observed nonlinear response is not solely governed by symmetry breaking, but is intimately linked to the emergence of coherent itinerant electronic states at low temperatures.  
These findings demonstrate that electronic functionalities in parity-violating AFM metals arise from the interplay between magnetic symmetry and electronic coherence.  

The parity-violating AFM state revealed here provides a new design principle for emergent electronic functionalities.  
Because global $\mathcal{P}$ breaking is intrinsically coupled to the AFM order in this class of matter, reversing the $\mathcal{T}$-related magnetic texture necessarily inverts $\mathcal{P}$-breaking responses.  
In other words, nonlinear electronic responses due to $\mathcal{P}$ breaking are intrinsically locked to the $\mathcal{T}$-related magnetic texture, and the $\mathcal{T}$-odd nonlinear susceptibility $\chi^{2\omega}_{\mathrm{AFM}}$ detected here provides a direct signature of this symmetry-locking.  
Such a symmetry-locked state is distinct from other systems where $\mathcal{PT}$ symmetry is broken~\cite{Ideue_nphys2017, Okuda2013, Dil2019}, as well as from recently proposed altermagnetic systems~\cite{jungwirth2026review}, and offers a deterministic route to control momentum-space asymmetry via manipulation of the magnetic texture.

Also, the metallic nature of LaMnSi is crucial in this context.  
While related symmetry-driven functionalities have predominantly been explored in insulating multiferroics~\cite{fiebig2005iop, tokura2014iop}, our results extend this concept to itinerant-electron systems.  
This not only enables direct access to nonreciprocal transport phenomena~\cite{tokura2018nonreciprocal}, but also suggests that current-driven control of AFM textures can simultaneously manipulate $\mathcal{P}$-breaking responses through the underlying symmetry-locking.  
The current-induced control of AFM state has indeed been demonstrated~\cite{wadley2016science,Bodnar2018,nair2020nmat}, indicating that a similar level of controllability is expected in the $\mathcal{P}$-breaking responses here.  
Therefore, our results establish parity-violating AFM metals as a universal materials platform for creating and controlling nonreciprocal and nonlinear electronic functionalities based on symmetry locking and coherent itinerant electrons.

\section*{\protect\NoCaseChange{Methods}}
\noindent\textbf{Sample growth and characterization.}

LaMnSi single crystals were prepared by a self-flux method using an alumina crucible.
Thin crystals were obtained with typical dimensions of 2$\times$2$\times$0.2~mm$^3$.
Details of the single-crystal growth and characterization of LaMnSi are described elsewhere~\cite{tanida2023jpsj}.\\

\noindent
\textbf{ARPES measurements.}

Soft x-ray ARPES measurements were performed at the BL25SU beamline of SPring-8 using a Scienta-Omicron DA30 electron analyzer~\cite{Muro2021jsr}.
The experiments utilized circularly polarized light with photon energies ($h\nu$) ranging from 400 to 600~eV.
The incident synchrotron radiation was focused to a spot size of less than 10~\textmu m at the sample surface.
The energy resolution was set to approximately 80~meV.

Vacuum ultraviolet ARPES experiments were performed at BL9A beamline of the Research Institute for Synchrotron Radiation Science in Hiroshima University (HiSOR) using a Scienta-Omicron R4000 electron analyzer.
These measurements utilized $p$-polarized light with a photon energy of $h\nu = 30$~eV.
The energy resolution was set to 20~meV.

In all of these ARPES experiments, the base pressure in the chamber was better than $2 \times 10^{-8}$~Pa.
The single crystals were cleaved \textit{in situ} at 60~K.
For the measurements of the temperature dependence, the ARPES spectra were sequentially acquired by scanning from lower to higher temperatures.

\noindent\\
\textbf{SHG microscopy.}

Polarization-resolved SHG microscopy measurements were performed using a mode-locked Yb-doped fiber laser (ORIGAMI, NKT Photonics) as the fundamental light source (center wavelength $\lambda = 1030$~nm, pulse duration $\sim 200$~fs, repetition rate 100~MHz). 
The detail of the setup is described in Supplementary Note~6.
To obtain a flat cleaved surface, the samples were cleaved in air and transferred to the vacuum cryostat within 3 minutes.
The laser beam was focused onto the cleaved (001) surface of the sample at an incidence angle of $45^\circ$ with a spot diameter of approximately 10 \textmu{m}.
The reflected SHG signal ($2\omega$, 515~nm) was spectrally isolated using appropriate short-pass and bandpass filters.
For high-sensitivity detection of minute signals, we employed a lock-in amplifier to measure the SHG signal from the biased photomultiplier tube (PMM01, Thorlabs), using the optical chopper frequency as a reference.
The polarization states of the fundamental ($\varphi_\omega$) and second-harmonic ($\varphi_{2\omega}$) electric fields were controlled using half-wave plates and nanoparticle linear film polarizers.
To visualize the magnetic domains, scanning SHG images were acquired by raster-scanning the sample using a stepper-motorized stage with a step size of $20$~\textmu m under fixed polarization configurations.
The sample was mounted in a liquid-helium-flow cryostat, and the measurements were conducted at temperatures ranging from 10~K to 300~K.

\noindent\\
\textbf{DFT calculations.}

Band calculations were performed using the DFT-based \textit{ab initio} calculation package WIEN2k (version 24.1)~\cite{blaha2020wien2k}, employing the generalized gradient approximation (GGA)~\cite{Perdew1996prl} for the exchange-correlation potential.
The muffin-tin radii ($R_{\rm MT}$) were set to 2.35 bohr for La and Mn, and 2.22 bohr for Si.
The plane-wave cutoff parameter $R_{\rm MT} K_{\max}$ was chosen to be 9.0, and a dense $k$-point mesh of $41 \times 41 \times 23$ was employed for the Brillouin zone integration.
The self-consistent cycles were converged until the energy and charge differences were less than $10^{-6}$ Ry and $10^{-5}$ $e$, respectively.
Spin-orbit coupling was included self-consistently to reproduce the band splitting and momentum-space asymmetry derived from the magnetic symmetry breaking.
The magnetic moments were initialized to form the $c$-axis collinear antiferromagnetic order consistent with the experimental magnetic structure.
The realistic tight-binding model for the AFM states of LaMnSi was constructed using WANNIER90~\cite{pizzi2020wannier90}.
The model was constructed by including Mn $3d$, La $4f$ and $5d$, and Si $3s$ and $3p$ orbitals as initial projections, and fitting the hopping parameters between them on an $8\times 8\times 6$ real-grid.
The disentanglement procedure was performed only within an energy window from $-10$~eV to $+4$~eV relative to the Fermi level, without further iterative wannierization.
The bulk band structures and Fermi surfaces presented in this study were then calculated by diagonalizing the derived Wannier Hamiltonian with dense $k$-meshes.

\section*{\protect\NoCaseChange{Acknowledgements}}
This work is supported by JSPS KAKENHI (Grants No. JP21H04652, No. JP23K17671, JP23K23211, JP23K20824, JP24K06943, JP24KK0107, JP24H01670, JP25H00743, JP25KJ1866), by JST FOREST Program (Grant No. JPMJFR236H and No. JPMJFR2444), by the Murata Science and Education Foundation, by Yamada Science Foundation, and also by the Collaborative Research Projects of Laboratory for Materials and Structures, Institute of Innovative Research, Science Tokyo.
T.I. acknowledges support from JST SPRING (Grant No. JPMJSP2132).
The SX-ARPES experiments were performed with the approval of Japan Synchrotron Radiation Research Institute (JASRI) (Proposals No. 2022A1434, 2022B1357, 2022A2060, 2022B2106, 2024A1625, 2024B1744, 2025A1614).
The VUV-ARPES measurements were performed with the approval of the proposal assessing committee of Research Institute for Synchrotron Radiation Science (Proposals No. 21BG012 and 21BG046)

\noindent\\
\noindent\textbf{Author contribution}\\
K.K. and H.T. planned the project.
T.I. and K.S. performed the ARPES experiments with D.S., T.T., Y.F., M.N., K.N., K.Y., M.A., A.K..
H.T. made and provided high-quality LaMnSi single crystals.
T.I. and T.A. performed polarizing SHG microscopy and analyzed the data.
T.Y. and Y.Y. calculated the band structures and provided theoretical insights.
T.I. and K.K. wrote the paper.
All authors discussed the results and commented on the manuscript.

\noindent\\
\textbf{Additional information}\\
The authors declare that they have no competing financial interests.\\	
\noindent\\
\textbf{Correspondence}\\
Correspondence and requests for materials should be addressed to Kenta~Kuroda~(email: kuroken224@hiroshima-u.ac.jp).

\bibliography{TI_LMS_v3}

\input{Supple_arXiv}

%

\end{document}

%% file: Supple_arXiv.tex






\clearpage 
\onecolumngrid 

\begin{center}
    \vspace*{2em}
    {\Large \bfseries Supplementary Information for:\\[0.5em]
    Realization of a parity-violating antiferromagnetic state in LaMnSi} 
    \vspace{2em}

    {\large
    Takuma~Iwata$^{1, 2}$, K.~Shiraishi$^{1}$, T.~Aoyama$^{1, 2}$, D.~Senba$^{1}$, T.~Takeda$^{1, 3}$, Y.~Fujisawa$^{4}$, M.~Nurmamat$^{1}$, K.~Nakanishi$^{1}$, K.~Yamagami$^{5}$, M.~Arita$^{4}$, T.~Yamada$^{6}$, Y.~Yanagi$^{6}$, A.~Kimura$^{1, 2, 7, 8}$, H.~Tanida$^{6}$, and Kenta~Kuroda$^{1, 2, 7}$
    } \\[1.5em] 

    {\small \itshape
    $^1$Graduate School of Advanced Science and Engineering, Hiroshima University, Higashi-Hiroshima 739-8526, Japan\\
    $^2$International Institute for Sustainability with Knotted Chiral Meta Matter (WPI-SKCM$^2$), Hiroshima University, Higashi-Hiroshima, 739-8531, Japan\\
    $^3$Department of Chemical System Engineering, Graduate School of Engineering, The University of Tokyo, Bunkyo-ku, Tokyo 113-8656, Japan\\
    $^4$Research Institute for Synchrotron Radiation Science, Hiroshima University, Higashi-Hiroshima 739-0046, Japan\\
    $^5$Japan Synchrotron Radiation Research Institute (JASRI), Sayo, Hyogo 679-5198, Japan\\
    $^6$Liberal Arts and Sciences, Toyama Prefectural University, Imizu, Toyama 939-0398, Japan\\
    $^7$Research Institute for Semiconductor Engineering, Hiroshima University, Higashi-Hiroshima 739-8527, Japan\\
    $^8$Synchrotron Radiation Research Center, National Institutes for Quantum Science and Technology (QST), Sayo-gun, Hyogo 679-5148, Japan\\
    }
    \vspace{2em}
\end{center}

\vspace{2em} 
\clearpage
\title{Realization of a parity-violating antiferromagnetic state in LaMnSi}
	\author{Takuma Iwata}
	\affiliation{Graduate School of Advanced Science and Engineering, Hiroshima University, Higashi-Hiroshima 739-8526, Japan}
	\affiliation{International Institute for Sustainability with Knotted Chiral Meta Matter (WPI-SKCM$^2$), Hiroshima University, Higashi-Hiroshima, 739-8531, Japan}
	
	\author{K.~Shiraishi} 
	\affiliation{Graduate School of Advanced Science and Engineering, Hiroshima University, Higashi-Hiroshima 739-8526, Japan}
	
	\author{T.~Aoyama} 
	\affiliation{Graduate School of Advanced Science and Engineering, Hiroshima University, Higashi-Hiroshima 739-8526, Japan}
	\affiliation{International Institute for Sustainability with Knotted Chiral Meta Matter (WPI-SKCM$^2$), Hiroshima University, Higashi-Hiroshima, 739-8531, Japan}
	
	\author{D.~Senba} 
	\affiliation{Graduate School of Advanced Science and Engineering, Hiroshima University, Higashi-Hiroshima 739-8526, Japan}
	
	\author{T.~Takeda}
    \affiliation{Graduate School of Advanced Science and Engineering, Hiroshima University, Higashi-Hiroshima 739-8526, Japan}
	\affiliation{Department of Chemical System Engineering, Graduate School of Engineering, The University of Tokyo, Bunkyo-ku, Tokyo 113-8656, Japan}
	
	\author{Y.~Fujisawa} 
	\affiliation{Research Institute for Synchrotron Radiation Science, Hiroshima University, Higashi-Hiroshima 739-0046, Japan}
	
	\author{M.~Nurmamat} 
	\affiliation{Graduate School of Advanced Science and Engineering, Hiroshima University, Higashi-Hiroshima 739-8526, Japan}
	
	\author{K.~Nakanishi} 
    \affiliation{Graduate School of Advanced Science and Engineering, Hiroshima University, Higashi-Hiroshima 739-8526, Japan}
	
	\author{K.~Yamagami} 
	\affiliation{Japan Synchrotron Radiation Research Institute (JASRI), Sayo, Hyogo 679-5198, Japan}

    \author{M.~Arita} 
	\affiliation{Research Institute for Synchrotron Radiation Science, Hiroshima University, Higashi-Hiroshima 739-0046, Japan}

    \author{T.~Yamada} 
	\affiliation{Liberal Arts and Sciences, Toyama Prefectural University, Imizu, Toyama 939-0398, Japan}
    
	\author{Y.~Yanagi} 
	\affiliation{Liberal Arts and Sciences, Toyama Prefectural University, Imizu, Toyama 939-0398, Japan}
	
	\author{A.~Kimura} 
	\affiliation{Graduate School of Advanced Science and Engineering, Hiroshima University, Higashi-Hiroshima 739-8526, Japan}
	\affiliation{International Institute for Sustainability with Knotted Chiral Meta Matter (WPI-SKCM$^2$), Hiroshima University, Higashi-Hiroshima, 739-8531, Japan}
	\affiliation{Research Institute for Semiconductor Engineering, Hiroshima University, Higashi-Hiroshima 739-8527, Japan}
	\affiliation{Synchrotron Radiation Research Center, National Institutes for Quantum Science and Technology (QST), Sayo-gun, Hyogo 679-5148, Japan}
	
	\author{H.~Tanida} 
	\affiliation{Liberal Arts and Sciences, Toyama Prefectural University, Imizu, Toyama 939-0398, Japan}
	
	\author{Kenta~Kuroda}
	\affiliation{Graduate School of Advanced Science and Engineering, Hiroshima University, Higashi-Hiroshima 739-8526, Japan}
	\affiliation{International Institute for Sustainability with Knotted Chiral Meta Matter (WPI-SKCM$^2$), Hiroshima University, Higashi-Hiroshima, 739-8531, Japan}
	\affiliation{Research Institute for Semiconductor Engineering, Hiroshima University, Higashi-Hiroshima 739-8527, Japan} 
\date{\today}
\maketitle
\subsection{DFT calculations of momentum-asymmetric electronic structures.}
\begin{suppfigure}[thpb]
\begin{center}
 \includegraphics{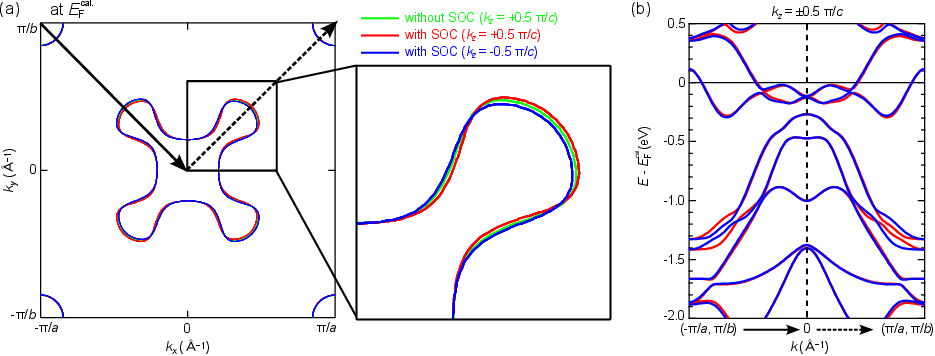}
 \caption{
 \textbf{Calculated Fermi surfaces and band dispersions in the AFM phase.} 
 (\textbf{a}) Calculated constant-energy contours at intermediate $k_z$ planes. 
 The green curves represent the symmetric contours obtained without SOC at $k_z = 0.5\pi/c$. 
 The red and blue curves denote the results with SOC at $k_z = +0.5\pi/c$ and $-0.5\pi/c$, respectively. 
(\textbf{b}) Calculated band dispersions with SOC along $(-\pi/a, +\pi/b)$-(0, 0)-$(+\pi/a, +\pi/b)$ cut, where a finite momentum-space asymmetry is expected [the solid and dashed arrows in (\textbf{a})].
The red and blue solid curves correspond to the bands at $k_z = +0.5\pi/c$ and $-0.5\pi/c$, respectively. 
The momentum asymmetry [$E(\mathbf{k}) \neq E(-\mathbf{k})$] clearly reverses sign upon switching the sign of $k_z$, corroborating the odd-parity nature of the magnetic quadrupole order based on magnetic point group.}
\label{fig1_s}
\end{center}
\end{suppfigure}
To elucidate the microscopic origin of the momentum-space asymmetry in LaMnSi, we analyzed the electronic bands using DFT calculations.
From a symmetry perspective, the antiferromagnetic (AFM) structure of LaMnSi belongs to the magnetic point group $4'/m'm'm$~\cite{tanida2022jpsj_s}.
To systematically link such macroscopic symmetry breaking to the underlying electronic structure and thereby predict emergent physical properties, the framework of magnetic multipoles provides an essential theoretical guide~\cite{hayami2024jpsj_s, Hayami2018prb_s, Suzuki2017prb_s}.
Indeed, this magnetic point group is shared by representative odd-parity magnetic multipole materials such as BaMn$_2$As$_2$ and EuMnBi$_2$~\cite{yanase2014_s, watanabe2017prb_s, shiomi2019prl_s}.
This magnetic order corresponds to the $B_{1u}$ irreducible representation of the tetragonal point group, which has the same symmetry properties as a magnetic quadrupole~\cite{watanabe2017prb_s}.
In momentum space, this symmetry breaking manifests as a functional term proportional to $k_x k_y k_z$.
Crucially, this term modulates the band dispersion only in the presence of spin-orbit coupling (SOC), thereby inducing a finite momentum-space asymmetry.

To investigate the impact of SOC on the Fermi surface distortion, we present the calculated Fermi surfaces with and without SOC in Supplementary Fig.~S\ref{fig1_s}(\textbf{a}).
The symmetric contour calculated without SOC (green) serves as a reference.
In contrast, incorporating SOC results in a distinct mismatch between the contours at $k_z = +0.5\pi/c$ (red) and $-0.5\pi/c$ (blue), demonstrating the asymmetric electronic structure with respect to $k_z=0$. 
Notably, the $k_x k_y k_z$ functional form implies that the distortion vanishes at high-symmetry planes (e.g., $k_z$=0) and becomes finite only at intermediate $k_z$ positions. 
We emphasize that this momentum dependence strictly necessitates $k_z$-resolved measurements, motivating our SX-ARPES experiments targeting the $k_z = \pm 0.5\pi/c$ planes, where the asymmetry is expected to be finite.

In Supplementary Fig.~S\ref{fig1_s}(\textbf{b}), our calculations also show that the energy difference between $+k_z$ and $-k_z$ bands due to the asymmetric electronic structure reaches a maximum of approximately 60 meV (with $\approx$ 20~meV near $E_\text{F}$).
This energy scale may point to the role of the SOC of the Mn $3d$ orbitals. 
\newpage
\subsection{Temperature evolution of the electronic band structure by ARPES}
\begin{suppfigure}[thbp]
\begin{center}
\includegraphics{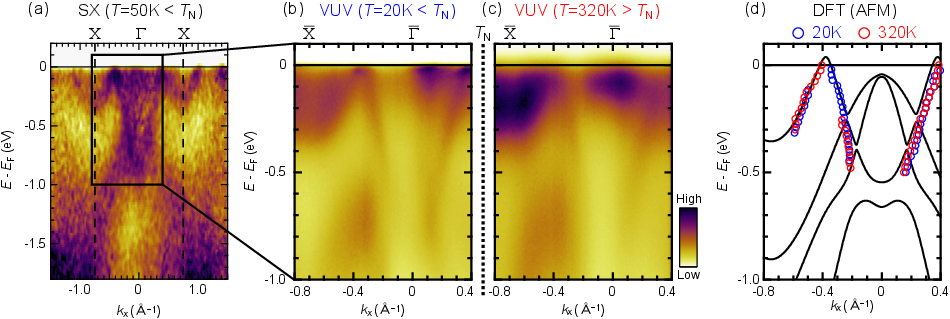}
\caption{
\textbf{Electronic structures across the N\'eel temperature.}
(\textbf{a}) SX-ARPES intensity map measured at 50~K along the $\Gamma$–X direction using a photon energy of $h\nu = 534$~eV [identical to Fig.~3(\textbf{e}) in the main text].
(\textbf{b}), (\textbf{c}) VUV-ARPES intensity maps measured with $h\nu = 30$~eV within the $E$-$k_x$ window shown in (\textbf{a}) at 20~K (\textbf{b}) and 320~K (\textbf{c}).
(\textbf{d}) Experimental peak positions extracted from momentum-distribution-curve analysis of (\textbf{b}) and (\textbf{c}), overlaid on the calculated AFM band structure along $\Gamma$–X line [identical to Fig.~3(\textbf{g}) in the main text].
}
\label{fig2_s}
\end{center}
\end{suppfigure}
To investigate the temperature evolution of the electronic structure across the N\'eel temperature ($T_{\rm N} = 293$~K), we performed temperature-dependent ARPES measurements using vacuum ultraviolet (VUV) light.
We first compare the VUV-ARPES results obtained at 20~K, well below $T_{\rm N}$ [Supplementary Fig.~S\ref{fig2_s}(\textbf{b})], with the bulk-sensitive SX-ARPES data measured at 50~K [Supplementary Fig.~S\ref{fig2_s}(\textbf{a}), identical to Fig.~3(\textbf{e}) in the main text].
The VUV-ARPES intensity map, measured within the energy–momentum window indicated by the solid box in Supplementary Fig.~S\ref{fig2_s}(\textbf{a}), is consistent with the SX-ARPES results [Supplementary Fig.~S\ref{fig2_s}(\textbf{b})].
Specifically, the VUV measurements capture the characteristic band dispersions consistent with the AFM calculations discussed in the main text, namely electron-like bands centered at the $\mathrm{X}$ and ${\Gamma}$ points.

We next examine the temperature evolution of these band dispersions across $T_{\rm N}$. The band structure shows no significant change with increasing temperature [Supplementary Figs.~S\ref{fig2_s}(\textbf{b})-(\textbf{d})].
Even at 320~K, well above $T_{\rm N}$ in the paramagnetic phase, the characteristic band features remain intact [Supplementary Fig.~S\ref{fig2_s}(\textbf{c})].
This behavior is further confirmed by comparing the peak positions extracted from momentum-distribution-curve (MDC) analysis [Supplementary Fig.~S\ref{fig2_s}(\textbf{d})].
The dispersions obtained at 20~K and 320~K nearly overlap and agree well with the calculated AFM band structure.
We note that the paramagnetic calculation without magnetic spin order fails to reproduce the electron-like band at the $\Gamma$ point [see Figs.~3(\textbf{g}) and 3(\textbf{h}) in the main text].
This indicates that the electron-like band associated with the AFM state persists even in the paramagnetic phase, implying the presence of robust short-range magnetic correlations.

\newpage
\subsection{Comparison of Fermi energy between ARPES and DFT.}
\begin{suppfigure}[thbp]
\begin{center}
\includegraphics{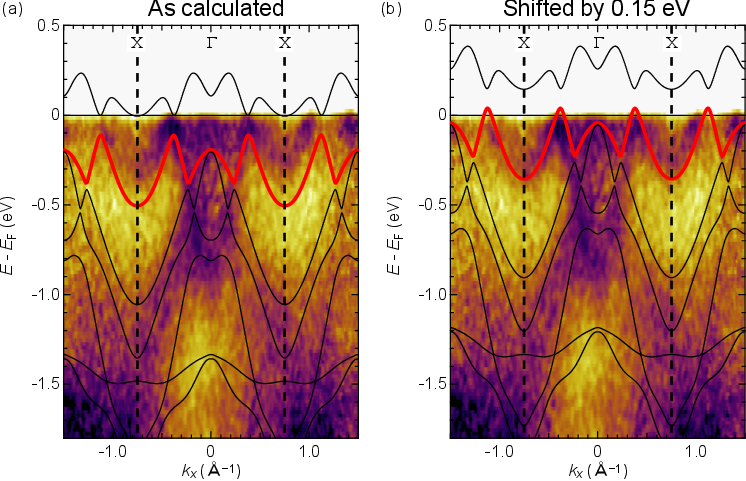}
\caption{
\textbf{Evaluation of the Fermi level alignment.}
(\textbf{a}) Superposition of the DFT band structure (solid curves) on the experimental ARPES intensity map along the $\Gamma$--X line.
The initial DFT calculation, without any energy shift, shows a discrepancy with the observed electronic states.
(\textbf{b}) Same as (\textbf{a}), but with the DFT bands rigidly shifted by 0.15~eV (solid curves). The shifted bands, particularly the band near $E_{\text{F}}$ indicated by the bold red curve, exhibit excellent agreement with the experimental ARPES features.}
\label{fig3_s}
\end{center}
\end{suppfigure}
To understand the observed electronic structure, we compared the experimental ARPES data with the DFT-calculated band structure. 
In Supplementary Fig.~S\ref{fig3_s}(\textbf{a}), the initial DFT bands (solid curves) are superimposed on the experimental data. 
We find discrepancies in the energy positions of the bands.
For instance, the calculated electron-like band at the X point is located at a lower energy than the experimental signal, which cannot explain the observed electron pocket.
In addition, the unshifted calculation cannot account for the spectral intensity observed in the immediate vicinity of $E_{\text{F}}$, as it misses the top of the band near the $\Gamma$ point.

To reconcile these discrepancies, we applied a rigid shift of 0.15~eV to the calculated bands in Supplementary Fig.~S\ref{fig3_s}(\textbf{b}).
The band highlighted by the bold red curve demonstrates a better correspondence with the main spectral peaks near $E_{\text{F}}$, capturing the observed electron pocket at X and the energy position of the band top near $\Gamma$ (see also Fig.~3 in the main text for a comparison between ARPES and DFT results).
\newpage
\subsection{Photoelectron intensity map at Fermi surfaces.}
\begin{suppfigure}[thbp]
\begin{center}
\includegraphics[width=0.8\columnwidth]{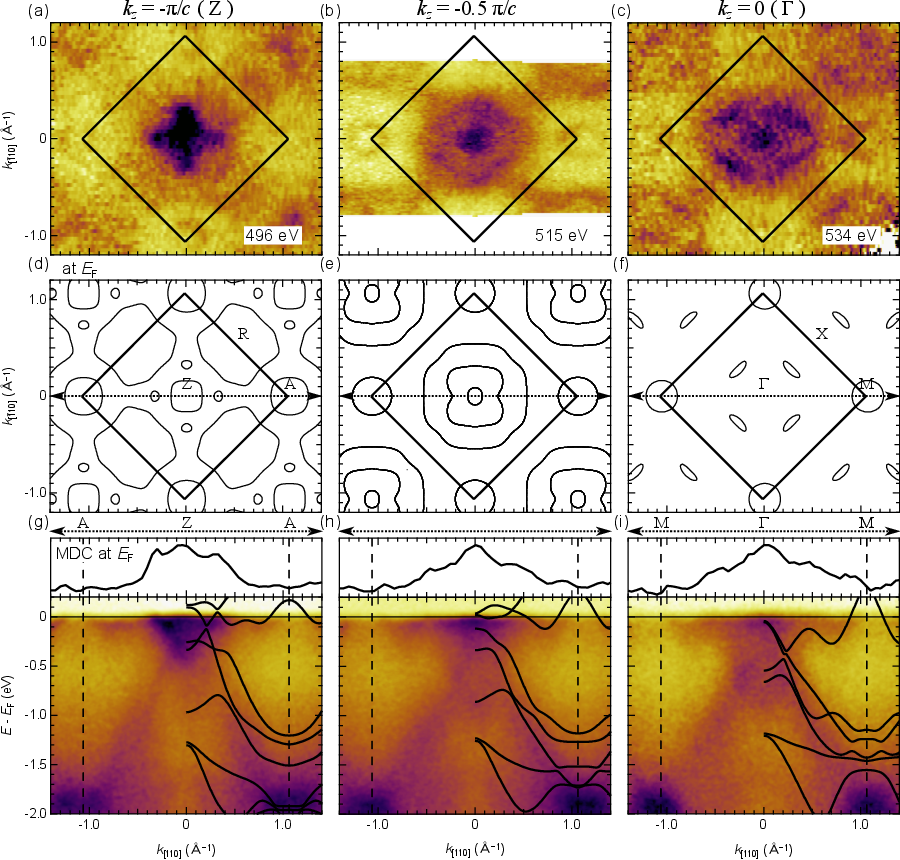}
\caption{
\textbf{Fermi-surface mappings at selected $k_z$ levels.}
(\textbf{a-c}) ARPES intensity maps at $E_\mathrm{F}$ measured at $k_z = -\pi/c$ (\textbf{a}), $-0.5\pi/c$ (\textbf{b}), and $0$ (\textbf{c}), using photon energies of 496, 515, and 534~eV, respectively. Solid square frames indicate the tetragonal Brillouin zone. 
(\textbf{d-f}) Corresponding calculated constant-energy contours at the experimental Fermi level.
(\textbf{g-i}) (Bottom): ARPES band maps along the cuts indicated by arrows in panels (\textbf{d-f}). Overlaid curves are theoretical bands. 
Dashed lines indicate Brillouin zone boundaries.
(Top): Corresponding MDCs at $E_\mathrm{F}$.}			
\label{fig4_s}
\end{center}
\end{suppfigure}
In the main text, we noted a discrepancy where significant photoelectron intensity is observed around the $\Gamma$ point in the experimental Fermi surface map [Supplementary Fig.~S\ref{fig4_s}(\textbf{c})], whereas the calculations predict no bands crossing $E_{\rm F}$ in this region [Supplementary Fig.~S\ref{fig4_s}(\textbf{f})].
This can be reconciled by examining the band dispersion in the immediate vicinity of $E_{\rm F}$.

As shown in the top panels of Supplementary Figs.~S\ref{fig4_s}(\textbf{g})-(\textbf{i}), the MDCs extracted at $E_{\rm F}$ clearly exhibit intensity peaks around the Brillouin zone center across all measured $k_z$ planes.
These experimental features are consistent with bands having relatively weak $k_z$ dispersion located just below $E_{\rm{F}}$, as shown in the theoretical band structures [Supplementary Figs.~S\ref{fig4_s}(\textbf{g})-(\textbf{i}), bottom panels].
Since the theoretical band maximum at the $\Gamma$ point lies slightly below $E_{\rm{F}}$, the observed intensity is naturally attributed to the spectral tail extending to $E_{\rm{F}}$.
This spectral tailing is caused by the finite experimental energy resolution likely combined with the finite $k_z$ broadening~\cite{Strocov2014jsr_s}.
\newpage
\subsection{Band character of Mn 3d orbitals}
\begin{suppfigure}[thbp]
\begin{center}
\includegraphics{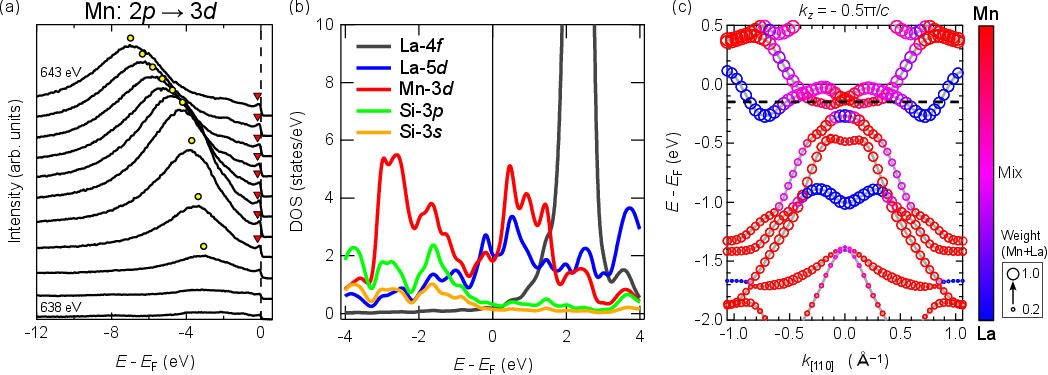}
\caption{
\textbf{Spectroscopic and theoretical evidence for the dual nature of Mn $3d$ electrons.}
(\textbf{a}) Mn $2p-3d$ resonant photoemission spectra ($h\nu = 638$--$643$~eV with a step of 0.5~eV).
(\textbf{b}) Calculated partial density of states (pDOS) showing the contribution of both Mn $3d$ and La $5d$ states near $E_{\text{F}}$.
(\textbf{c}) Orbital-projected band dispersion at $k_z = -0.5\pi/c$ along the $[110]$ direction.
The bands are weighted by the Mn $3d$ character, highlighting the strong correlation between the magnetic element and the conduction electrons responsible for the electronic asymmetry.
}
\label{fig5_s}
\end{center}
\end{suppfigure}
To experimentally elucidate the contributions of the Mn $3d$ orbitals in the observed band structures, we performed resonant photoemission spectroscopy (RPES)~\cite{hufner2013book_s}.
In Supplementary Fig.~S\ref{fig5_s}(\textbf{a}), we present the resonant photoemission spectra around the Mn $2p-3d$ core absorption region ($h\nu=638-643$~eV).
As $h\nu$ is tuned through the absorption edge, a characteristic peak in the obtained spectra develops around ${E}- E_{\text{F}}{\approx}-$3~eV and shifts toward higher binding energies with increasing $h\nu$ (indicated by the yellow circles).
This peak is characterized by a constant kinetic energy, identifying it as normal Auger emission.
In the RPES process, normal Auger emission occurs when the photoexcited orbitals possess itinerant character and delocalize before the core-hole decay, causing another valence electron to fill the core hole instead.
Therefore, this Auger feature directly reflects the itinerant nature of the Mn $3d$ electrons~\cite{weinelt1997prl_s, kono2019prb_s}.

We also note that an enhancement of the spectral intensity is observed in the immediate vicinity of $E_{\text{F}}$ (red triangles). 
Unlike the normal Auger peak, this enhancement occurs at a constant binding energy. 
This likely indicates a localized-like resonant process, where the photoexcited electron remains spatially localized and directly participates in the core-hole decay.
This coexistence of a localized-like resonance around $E_{\text{F}}$ and an itinerant normal Auger emission provides a spectroscopic signature of the dual nature of the Mn $3d$ electrons.

These features are further supported by the calculated partial density of states (pDOS) in Supplementary Fig.~S\ref{fig5_s}(\textbf{b}), which shows that the states near $E_{\text{F}}$ mainly exhibit Mn $3d$ character, with contributions from La $5d$-derived states.
We also show the orbital-projected band structure for the $k_z$=$-$0.5$\pi/c$ plane [Supplementary Fig.~S\ref{fig5_s}(\textbf{c})], corresponding to the dispersions [Figs.~4(\textbf{b}), (\textbf{e}) in the main text and Supplementary Fig.~S\ref{fig4_s}(\textbf{h})]. 
Intriguingly, a comparison between the orbital-projected band dispersions [Supplementary Fig.~S\ref{fig5_s}(\textbf{c})] and the asymmetric electronic structure [Supplementary Fig.~S\ref{fig1_s}(\textbf{b})] suggests that the electronic asymmetry is enhanced in the momentum regions where the Mn $3d$ orbital contribution is dominant.
\newpage
\subsection{Schematic picture of polarization-resolved SHG experiment setup}
\begin{suppfigure}[thbp]
\begin{center}
\includegraphics{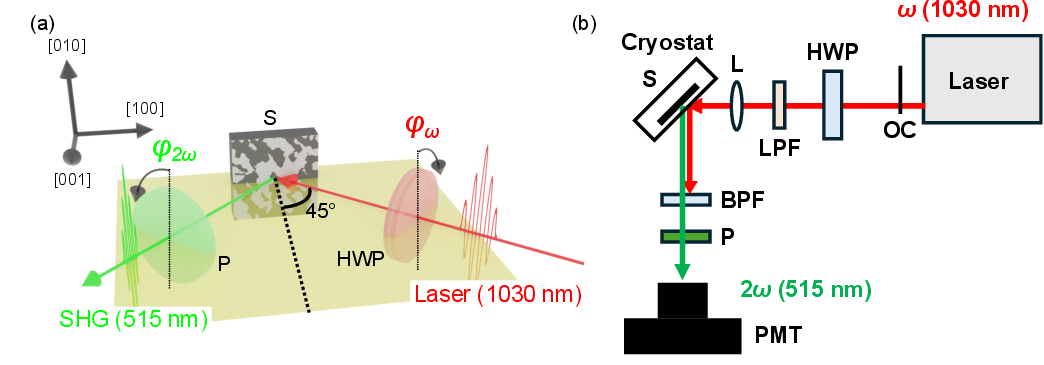}
\caption{
\textbf{Schematic of the SHG microscopy setup.}
(\textbf{a}) Definition of the polarization geometry and sample orientation. 
The crystallographic axes $a$ ([100]) and $b$ ([010]) lie in the sample surface plane. 
$\varphi_\omega$ and $\varphi_{2\omega}$ denote the polarization angles of the fundamental (red arrow) and second-harmonic (green arrow) electric fields, measured from the $b$ axis.
The yellow plane represents the plane of incidence.
(\textbf{b}) Detailed diagram of the optical setup. 
The system consists of a mode-locked Yb fiber laser, an optical chopper (OC), a half-wave plate (HWP) mounted on a rotation stage, a focusing lens (L), spectral filters [long-pass filter (LPF), band-pass filter (BPF)], a polarization analyzer (P) mounted on a rotation stage, and a photomultiplier tube (PMT) connected to a lock-in amplifier. 
The sample is mounted inside a cryostat, which is placed on a motorized stage to enable spatial mapping.}
\label{fig6_s}
\end{center}
\end{suppfigure}
Supplementary Figure~S\ref{fig6_s} illustrates the experimental geometry and the optical setup employed for the polarization-resolved second-harmonic generation (SHG) measurements~\cite{petersen2006nphys_s}.
As shown in Supplementary Fig.~S\ref{fig6_s}(\textbf{a}), the measurements were performed on the cleaved (001) surface of LaMnSi. 
The polarization angle of the fundamental light, $\varphi_{\omega}$, and that of the second-harmonic light, $\varphi_{2\omega}$, are defined relative to the $b$ axis (which is perpendicular to the plane of incidence).
Supplementary Fig.~S\ref{fig6_s}(\textbf{b}) shows the schematic diagram of the optical setup. 
The fundamental light source is a mode-locked Yb-doped fiber laser (center wavelength $\lambda = 1030$~nm, pulse duration $\sim 200$~fs, repetition rate 100~MHz). 
The laser beam is modulated by an optical chopper for lock-in detection. 
The polarization state of the incident fundamental beam is controlled by a rotating half-wave plate (HWP).
The beam is focused onto the sample surface to a spot size of approximately 10~\textmu m using a focusing lens (L), with an incident pulse energy of 2~nJ. 
The generated SHG signal (515~nm) is collected in the reflection geometry. 
The SHG signal is spectrally isolated by a band-pass filter (BPF), and its polarization is analyzed by a rotating wire-grid polarizer (P).
Finally, the signal is detected by a photomultiplier tube (PMT) and measured using a lock-in amplifier referenced to the chopper frequency.
The sample is housed in a liquid-helium-flow cryostat to control the temperature down to 10~K. 
For spatially resolved measurements (scanning SHG microscopy), the cryostat is mounted on a motorized two-axis stage, enabling two-dimensional scanning of the sample position relative to the fixed laser beam.
\newpage
\subsection{Theoretical derivation of the SHG intensity and tensor analysis}
\begin{suppfigure}[thbp]
\begin{center}
\includegraphics{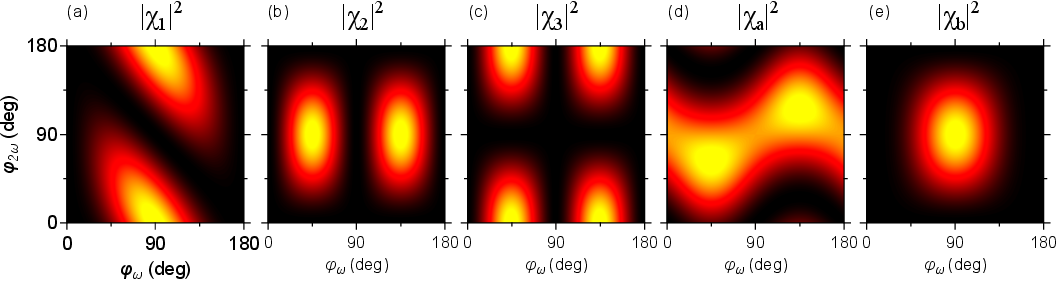}
\caption{
\textbf{Calculated polarization dependence of SHG intensity for individual susceptibility components.}
(\textbf{a})-(\textbf{e}) Simulated 2D intensity maps of the SHG response obtained using Eq.~\eqref{equ_Int}, plotted as a function of the fundamental polarization angle ($\varphi_{\omega}$) and the SHG polarization angle ($\varphi_{2\omega}$).
Each panel corresponds to the contribution from a single independent tensor component:
(\textbf{a}) $\chi_1$ (AFM),
(\textbf{b}) $\chi_2$ (AFM),
(\textbf{c}) $\chi_3$ (crystallographic),
(\textbf{d}) $\chi_a$ (surface), and
(\textbf{e}) $\chi_b$ (surface).
The color scale represents the normalized SHG intensity.
}
\label{fig7_s}
\end{center}
\end{suppfigure}
In this section, we derive the theoretical expression for the SHG intensity used to analyze the experimental results.
Optical SHG has long been established as a powerful probe of symmetry breaking.
Conventionally, the SHG response is primarily described within the electric-dipole (ED) approximation, where the induced electric polarization $P_i(2\omega)$ is proportional to the square of the fundamental electric field $E(\omega)$ via the third-rank nonlinear susceptibility tensor $\chi_{ijk}$, i.e., $P_i(2\omega) \propto \sum_{jk}\chi_{ijk} E_j(\omega) E_k(\omega)$ (where $i, j, k$ denote the crystallographic axes).
This standard mechanism has been extensively utilized to study systems with broken space-inversion ($\mathcal{P}$) symmetry, such as polar crystals and surfaces/interfaces.
Furthermore, because the allowed components of this macroscopic tensor $\chi_{ijk}$ are rigorously constrained by the point group symmetries, its functional form establishes a one-to-one correspondence with the underlying physical properties and symmetry-breaking order of the material~\cite{boyd2008book_s}.

However, this conventional ED approximation alone is often insufficient to fully capture the nonlinear optical responses in magnetic systems. 
To explicitly probe the breaking of time-reversal ($\mathcal{T}$) symmetry and completely distinguish complex magnetic structures, the formalism must be generalized by incorporating magnetic-dipole (MD) transitions. 
This is achieved by introducing source terms from the magnetization $\bm{M}(2\omega)$ induced by the fundamental electromagnetic field, in addition to the conventional electric polarization $\bm{P}(2\omega)$---an approach pioneered by the observation of AFM-induced SHG in the magnetoelectric insulator $\mathrm{Cr_2O_3}$~\cite{fiebig1994prl_s, fiebig2005josab_s}. 
Within this generalized framework, the macroscopic responses are described by two coupled equations governing $\bm{P}(2\omega)$ and $\bm{M}(2\omega)$.
Consequently, the associated nonlinear susceptibility tensors $\chi_{ijk}$ are classified into four distinct types based on their parity under $\mathcal{P}$ and $\mathcal{T}$ operations~\cite{fiebig2005josab_s}:
polar (odd under $\mathcal{P}$) vs. axial (even under $\mathcal{P}$); and $i$-type (even under $\mathcal{T}$) vs. $c$-type (odd under $\mathcal{T}$).
These relationships are comprehensively expressed as:
\begin{equation} 
\begin{pmatrix}
P_i(2\omega) \\[1.3ex]
M_i(2\omega) 
\end{pmatrix} 
= \varepsilon_0 \sum_{jk} 
\begin{pmatrix} 
\chi_{ijk}^{\text{polar}, i} & \chi_{ijk}^{\text{polar}, c} \\[1.3ex]
\chi_{ijk}^{\text{axial}, i} & \chi_{ijk}^{\text{axial}, c} 
\end{pmatrix} 
\begin{pmatrix} E_j(\omega) E_k(\omega) \\[1.3ex]
E_j(\omega) \dot{E}_k(\omega) 
\end{pmatrix} 
\label{eq1} 
\end{equation} 
where $\chi_{ijk}^{\text{polar}}$ and $\chi_{ijk}^{\text{axial}}$ represent the nonlinear susceptibility tensors that transform as polar and axial tensors, respectively.

In LaMnSi, the parity-violating AFM order macroscopically breaks $\mathcal{P}$ symmetry, thereby inducing the polar $c$-tensor, which changes sign under the $\mathcal{T}$ operation and serves as the primary focus of this study. 
Alongside this magnetic contribution, the system also allows responses that are invariant under $\mathcal{T}$ (i.e., even under $\mathcal{T}$): namely, the polar $i$-tensor originating from surface symmetry breaking and the axial $i$-tensor arising from the bulk crystallographic structure.
By applying the rigorous symmetry constraints of the $4'/m'm'm$ magnetic point group for the AFM phase of LaMnSi, we can explicitly determine the non-zero components of the SHG susceptibility tensor~\cite{birss1964symmetry_s}.
Here, the independent parameters are defined based on their physical origins and symmetry properties:
\begin{itemize}
\item AFM (polar-$c$ tensor):
$\chi_1 = \chi_{xyz} = \chi_{yzx}$ and $\chi_2 = \chi_{zxy}$.
\item Crystallographic (axial-$i$ tensor):
$\chi_3 = \chi_{xyz} = -\chi_{yzx}$.
\item Surface (polar-$i$ tensor):
$\chi_a = \chi_{xzx} = \chi_{yyz} = \chi_{zyy} = \chi_{zxx}$ and $\chi_b = \chi_{zzz}$.
\end{itemize}
Here, the Cartesian coordinates $x, y$, and $z$ correspond to the crystallographic $a, b$, and $c$ axes, respectively. 
All other components are constrained to be zero by the $4'/m'm'm$ symmetry.

Based on the non-zero tensor components, the AFM-induced nonlinear polarization $\bm{P}(2\omega)$ can be explicitly expanded using the standard $3 \times 6$ matrix notation:
\begin{equation}
\begin{pmatrix} 
 P_x \\
 P_y \\
 P_z 
\end{pmatrix}
\propto
\begin{pmatrix}
0 & 0 & 0 & \chi_1 & 0 & 0 \\
0 & 0 & 0 & 0 & \chi_1 & 0 \\
0 & 0 & 0 & 0 & 0 & \chi_2
\end{pmatrix}
\begin{pmatrix} 
E_x^2 \\[1ex]
E_y^2 \\[1ex]
E_z^2 \\[1ex]
2E_y E_z \\[1ex]
2E_z E_x \\[1ex]
2E_x E_y 
\end{pmatrix}
= 2\begin{pmatrix}
\chi_1 E_y E_z \\
\chi_1 E_z E_x \\
\chi_2 E_x E_y
\end{pmatrix}   
\end{equation}
This matrix representation explicitly shows that the generation of the in-plane polarizations ($P_x$ and $P_y$) strictly requires a $z$-component of the fundamental electric field ($E_z \neq 0$).
Under normal incidence on the (001) plane, the fundamental electric field is confined to the $xy$-plane ($E_z = 0$), which yields zero detectable AFM-induced SHG signal.
To introduce the necessary $E_z$ component and activate these tensor elements, we therefore adopt an oblique incidence geometry, as shown in Supplementary Fig.~S\ref{fig6_s}.

In our experimental geometry, the fundamental light is obliquely incident on the (001) surface.
The electric field components $E_i(\omega)$ are determined by the fundamental polarization angle $\phi_{\omega}$.
From Maxwell's equations, the radiation source term $\bm{S}(2\omega)$ is given by the time derivative of the nonlinear current density~\cite{fiebig1994prl_s}, and is explicitly expressed using the vacuum permeability $\mu_0$ as:
\begin{equation}
\bm{S}(2\omega) = \mu_0 \left( \frac{\partial^2 \bm{P}(2\omega)}{\partial t^2} + \nabla \times \frac{\partial \bm{M}(2\omega)}{\partial t} \right).
\label{eq:Source_Maxwell}
\end{equation}
By substituting the specific tensor elements into Eq.~\eqref{eq1} and projecting the vector $\bm{S}$ onto the analyzer polarization state defined by $\phi_{2\omega}$, we obtain the scalar effective source term $S(\phi_\omega, \phi_{2\omega})$.
Consequently, the detected SHG intensity $I_{\mathrm{SHG}}$, which is proportional to the square modulus of $S$, is given by:
\begin{equation}
\begin{aligned}
I_{\mathrm{SHG}}(\phi_\omega, \phi_{2\omega}) \propto |S(\phi_\omega, \phi_{2\omega})|^2 \propto
&~\Big|~\chi_1 \left\{ \sin(2\phi_\omega) \sin(\phi_{2\omega}) + 2\sin^2(\phi_\omega) \cos(\phi_{2\omega}) \right\} \\
&+ \chi_2 \sin(2\phi_\omega) \sin(\phi_{2\omega}) \\
&-\chi_3 \sin(2\phi_\omega) \cos(\phi_{2\omega}) \\
&- \sqrt{2}\chi_a \left\{ 2\sin(\phi_{2\omega}) + \sin(2\phi_\omega) \cos(\phi_{2\omega}) \right\} \\
&- \chi_b \sin^2(\phi_\omega) \sin(\phi_{2\omega})~\Big|^2.
\end{aligned}
\label{equ_Int}
\end{equation}
Using this formula, we fitted the experimental data shown in Figs.~5(\textbf{a2}) and (\textbf{b2}) in the main text.
The overall fitting results and the quantitative evaluation of the domain contrast are summarized in Supplementary Note 8.

For the analysis of the temperature dependence shown in Figs.~5(\textbf{a1}) and (\textbf{b1}) in the main text, we utilized a specific polarization geometry where the analyzer angle is fixed to $\phi_{2\omega} = 90^\circ$ ($p$-polarized detection).
By substituting $\phi_{2\omega} = 90^\circ$ into Eq.~\eqref{equ_Int}, the terms containing $\cos(\phi_{2\omega})$ vanish, and thus $\chi_3$ does not contribute in this specific geometry.
Under this geometry, the two AFM components $\chi_1$ and $\chi_2$ exhibit identical angular dependence. Thus, we define the AFM-induced susceptibility $\chi^{2\omega}_{\text{AFM}}$ as:
\begin{equation}
\chi^{2\omega}_{\text{AFM}} \equiv \chi_1 + \chi_2.
\end{equation}
Using this definition, the SHG intensity expression simplifies to:
\begin{equation}
I_{\mathrm{SHG}}(\phi_\omega, 90^\circ) \propto | \chi^{2\omega}_{\text{AFM}} \sin(2\phi_\omega) - 2\sqrt{2}\chi_a - \chi_b \sin^2(\phi_\omega) |^2.
\label{eq:Int_90deg}
\end{equation}
We used this simplified equation to fit the rotational anisotropy patterns obtained at various temperatures and extracted the AFM-induced susceptibility $\chi^{2\omega}_{\text{AFM}}$ in the main text.

\newpage
\subsection{Quantitative determination of the nonlinear susceptibility tensor components}
\begin{suppfigure}[thbp]
\begin{center}
\includegraphics{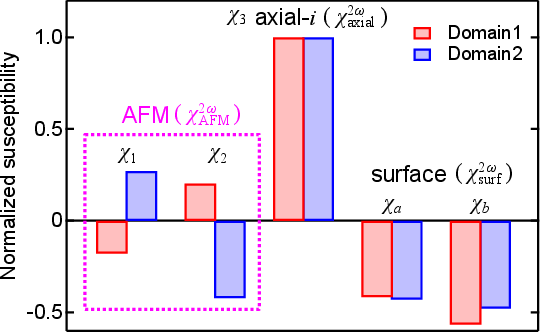}
\caption{
\textbf{Comparison of the extracted nonlinear susceptibility components between opposite magnetic domains.}
Bar graph showing the values of the five independent tensor components ($\chi_1, \chi_2, \chi_3, \chi_a, \chi_b$) obtained by fitting the SHG patterns in Figs.~5(\textbf{a2}) and (\textbf{b2}) in the main text.
The red and blue bars correspond to domain~1 and domain~2, respectively.
All values are normalized to the magnitude of the dominant crystallographic component, $\chi_3$.
Note that the polar-$c$ terms $\chi_1$ and $\chi_2$ corresponding to $\chi^{2\omega}_{\text{AFM}}$ reverse their signs between the domains, whereas the crystallographic and surface terms ($\chi_3, \chi_a, \chi_b$) remain invariant.
}
\label{fig8_s}
\end{center}
\end{suppfigure}
\begin{table}[h]
\centering
\caption{
\textbf{Summary of the fitted nonlinear susceptibility tensor components.} 
The values are normalized relative to the largest component ($\chi_3$). 
The uncertainties quoted correspond to the standard errors estimated from the least-squares fitting procedure.
}
\begin{tabular}{lcccc}
\hline \hline
Component & Origin & Symmetry & Domain 1 & Domain 2 \\ 
\hline
$\chi_1$ & AFM & $\mathcal{T}$-odd & $-0.18 \pm 0.02$ & $+0.27 \pm 0.02$ \\ [1.3ex]
$\chi_2$ & AFM & $\mathcal{T}$-odd & $+0.20 \pm 0.03$ & $-0.42 \pm 0.03$ \\ [1.3ex]
$\chi_3$ & Crystallographic & $\mathcal{T}$-even & $+1.00$ (fixed) & $+1.00$ (fixed) \\ [1.3ex]
$\chi_a$ & Surface & $\mathcal{T}$-even & $-0.42 \pm 0.01$ & $-0.43 \pm 0.01$ \\ [1.3ex]
$\chi_b$ & Surface & $\mathcal{T}$-even & $-0.57 \pm 0.06$ & $-0.48 \pm 0.05$ \\ [1.3ex]
\hline \hline
\end{tabular}

\label{tab:fitting}
\end{table}
In this section, we present the quantitative results of the tensor analysis described in Supplementary Note~7.
The experimental SHG polarization patterns for domain~1 and domain~2 [Figs.~5(\textbf{a2}), (\textbf{b2}) in the main text] were fitted using Eq.~(\ref{equ_Int}) to extract the magnitude and relative sign of the five independent susceptibility components ($\chi_1, \chi_2, \chi_3, \chi_a, \chi_b$).
The extracted best-fit parameters are summarized in Supplementary Table~\ref{tab:fitting} and visualized as a bar graph in Supplementary Fig.~S\ref{fig8_s}.

As derived in the previous section, the tensor components are classified based on their behavior under the $\mathcal{T}$ operation.
The AFM terms $\chi^{2\omega}_{\text{AFM}}$ ($\chi_1$ and $\chi_2$) are $\mathcal{T}$-odd, while the crystallographic $\chi^{2\omega}_{\text{axial}}$ ($\chi_3$) and surface $\chi^{2\omega}_{\text{surf}}$ ($\chi_a, \chi_b$) terms are $\mathcal{T}$-even~\cite{fiebig2005josab_s}.
The results in Supplementary Fig.~S\ref{fig8_s} clearly demonstrate this symmetry property based on the magnetic point group of LaMnSi.
The crystallographic and surface components exhibit constant values across the two AFM domains within experimental error.
In sharp contrast, the AFM components $\chi^{2\omega}_{\text{AFM}}$ ($\chi_1$ and $\chi_2$) show a clear sign reversal between domain~1 and domain~2.
This specific sign reversal is the hallmark of the $\mathcal{T}$ breaking associated with the parity-violating AFM order~\cite{fiebig2005josab_s, sun2019nature_s}.
The obtained fitting errors (Table~\ref{tab:fitting}) are sufficiently small compared to the magnitude of each component, ensuring the reliability of our quantitative analysis. 

We note that while the AFM components $\chi^{2\omega}_{\text{AFM}}$ ($\chi_1$ and $\chi_2$) exhibit clear sign reversals between the parity-violating AFM domains, their absolute magnitudes show slight deviations between domain~1 and domain~2. 
These minor asymmetries are likely attributed to experimental factors rather than an intrinsic physical asymmetry; specifically, slight misalignment in the $45^\circ$ reflection geometry arising from variations in the cleaved surfaces, as well as subtle differences in surface conditions between the measured domain areas. 
Such extrinsic factors can influence the effective projection of the nonlinear susceptibility components.
Crucially, the disentanglement of such sign-reversing terms from the invariant background allows for a clear determination of the magnetic contribution.
These quantitative analyses provide definitive evidence that the observed SHG response intrinsically reflects the parity-violating AFM state of LaMnSi.